\documentstyle[twocolumn,aps,epsfig,floats]{revtex}
\draft 
\begin{document} 
\twocolumn[\hsize\textwidth\columnwidth\hsize\csname @twocolumnfalse\endcsname
\title{ Itinerant ferromagnetism in half-metallic CoS$_2$} 
\author{ S. K. Kwon, S. J. Youn\cite{AAAuth}, and B. I. Min } 
\address{ Department of Physics, 
          Pohang University of Science and Technology, 
          Pohang 790-784, Korea } 
\date{\today}
 
\maketitle 
 
\begin{abstract}    
We have investigated 
electronic and magnetic properties of the pyrite-type CoS$_2$ 
using the linearized muffin-tin orbital (LMTO) band method. 
We have obtained the ferromagnetic ground state with nearly half-metallic 
nature.  The half-metallic stability is studied by using the fixed 
spin moment method.
The non-negligible orbital magnetic moment of Co $3d$ electrons 
is obtained as $\mu_L = 0.06 \mu_B$ in the local spin density 
approximation (LSDA).
The calculated ratio of the orbital to spin angular momenta 
$\langle L_z \rangle/\langle S_z \rangle = 0.15$
is consistent with experiment.
The effect of the Coulomb correlation between Co $3d$ electrons is 
also explored with the LSDA + $U$ method. 
The Coulomb correlation at Co sites is not so large, $U \lesssim 1$ eV,
and so CoS$_2$ is possibly categorized as an itinerant ferromagnet. 
It is found that the observed electronic and magnetic 
behaviors of CoS$_2$ can be described better by the LSDA 
than by the LSDA + $U$. 
\end{abstract}

%
]
\narrowtext  
Transition metal disulfides of MS$_2$ (M = Fe, Co, Ni, etc.) 
with the pyrite structure shows a variety of physical properties.
Electronic and magnetic properties of the MS$_2$ has been 
qualitatively explained by the successive filling of the $e_g$ band 
of metal ions.  In the NaCl-type structure of pyrite MS$_2$, 
divalent metal ions of M$^{2+}$ are located 
at the center of octahedron formed by S$_2$$^{2-}$ dimers.
Hence, the M-$3d$ bands are split into $t_{2g}$ and $e_g$
subbands due to large crystal field.
The M-$3d$ electron configuration varies from $t_{2g}^6e_g^0$,
$t_{2g}^6e_g^1$, to $t_{2g}^6e_g^2$ for FeS$_2$, CoS$_2$, and NiS$_2$.
All of them are in the low-spin states, $S=0$, $S=1/2$, and $S=1$ for
FeS$_2$, CoS$_2$, and NiS$_2$, respectively.
With fully occupied $t_{2g}^6$ and completely empty $e_g^0$ bands,
FeS$_2$ is a semiconductor with an energy gap of $E_g \simeq 0.8$ eV\cite{Li}.
CoS$_2$ is a ferromagnetic metal with $T_C \simeq 120$ K\cite{Andresen,Adachi1}.
NiS$_2$ is an antiferromagnetic insulator ($T_N \simeq 40$ K)
with a  half-filled $e_g^2$ band due to the large
on-site Coulomb correlation\cite{Wilson,Fujimori}. 

Despite extensive studies on these systems, 
the electronic and magnetic properties of CoS$_2$ are not
fully understood yet. 
The issues for CoS$_2$ to be resolved are $(i)$ whether it is a strongly 
correlated electron system, $(ii)$ whether it is classified as an
itinerant electron ferromagnet,
and $(iii)$ whether it is a half-metal or not.
It is considered that Co $3d$ states in CoS$_2$ are just
at the boundary of the localized and delocalized regime\cite{Wilson}.
Magnetic measurement using the polarized neutron diffraction indicates 
that most part of the magnetic moment is localized at Co sites\cite{Ohsawa}.
The analysis of photoemission spectroscopy (PES) data for CoS$_2$
in terms of the cluster model calculations yields a large Coulomb 
correlation parameter, $U = 3.0 \sim 4.2$ eV\cite{Fujimori,Muro}.
According to the magnetic circular dichroism (MCD) measurements,
a non-negligible orbital magnetic moment is observed for Co $3d$ electrons, 
$\langle L_z \rangle/\langle S_z \rangle = 0.18$\cite{Muro}
and $0.14$\cite{Miyauchi}.
Such an unquenched orbital moment is exceptional, 
considering the low-spin state of Co $3d$ electrons in CoS$_2$    
with a large crystal field splitting of $10D_q \simeq 2.5$ eV. 
On the other hand, transport properties of CoS$_2$ are known
to be well described by the itinerant electron model\cite{Jarrett,Adachi2}.
Further, the dynamical susceptibility measured by the
inelastic magnetic neutron scattering\cite{Hiraka} and 
optical properties\cite{Yamamoto} are analyzed 
based on the itinerant band model.

Using the linearized atomic orbitals (LCAO) band method,
Zhao {\it et al.}\cite{Zhao} has found that CoS$_2$ is
close to half-metallic 
in the sense that the occupation number of the minority spin $e_g$ band is   
very small compared to that of the majority spin $e_g$ band.
In fact, recent experimental reflectivity data have been consistently 
interpreted under the assumption of half-metallic CoS$_2$\cite{Yamamoto}.
In contrast, the  linearized muffin-tin orbital (LMTO)
band calculation by Yamada {\it et al.}\cite{Yamada} gives
rise to the normal-metallic ferromagnetic ground state with
a partially filled minority spin $e_g$ band.

To clarify the above mentioned issues and
to resolve the differences between existing band calculations,
we have reexamined electronic and magnetic properties of CoS$_2$ 
using the LMTO band method.
Both the local spin density approximation
(LSDA) and the LSDA + $U$ approximation\cite{Anisimov} are employed
to explore the effect of the Coulomb correlation interaction 
between Co $3d$ electrons.  The von Barth-Hedin form of
the exchange-correlation potential is utilized.
Further, the fixed spin moment method is used 
to check the stability of half-metallic state.
We have also estimated the orbital contribution to the total magnetic 
moment in CoS$_2$, using the fully relativistic calculations 
in which the spin-orbit coupling is simultaneously included 
in the self-consistent variational loop\cite{Min}. 
Two main parameters in the LSDA + $U$ method are the Coulomb $U$ 
and the exchange $J$ interactions.
The relations of these parameters to the Slater integrals 
are given by $U = F^{0}$ and $J = (F^{2}+F^{4})/14$. 
The ratio of $F^{4}/F^{2}$ is known to be constant 
around 0.625 for most $3d$ transition metal atoms\cite{Sawatzky}.  
We have used parameter values of $U = 0.0 \sim 4.0$ eV 
with a fixed $J = 0.89$ eV.

The crystal structure of pyrite CoS$_2$ is cubic with
the lattice constant of $a = 5.528$ \AA~ 
and the space group is $T_h^6 (Pa\bar3)$.
There are twenty-four symmetry operations.
The four Co atoms are located at positions $(4a)$ : 
(0,0,0), (1/2,0,1/2), (0,1/2,1/2), and (1/2,1/2,0). 
The eight S atoms are at positions $(8c)$ : 
$(u,u,u)$, $(u+1/2,u,\bar{u}+1/2)$, $(u,\bar{u}+1/2,u+1/2)$, 
and $(\bar{u}+1/2,u+1/2,u)$ where we have used the position parameter 
of $u = \pm 0.389$.
To improve the packing ratio, 
twenty-four empty spheres are introduced at positions $(24d)$.
For the Brillouin zone (BZ) integration, the tetrahedron method is adopted 
with 80 k-points sampling in the irreducible BZ wedge.
The LMTO basis functions are included up to $l = 2$ for Co, 
and $l = 1$ for S and empty spheres. 

\begin{figure}[t]
\epsfig{file=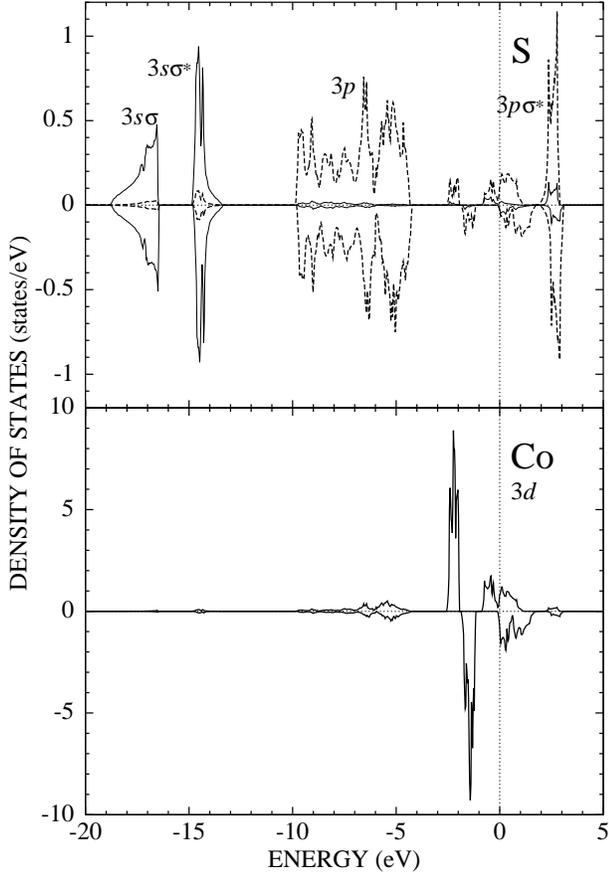,width=8.65cm}
\caption{\label{fig1} The LSDA DOS of pyrite CoS$_2$ at each atomic site.
The two peaks at the highest binding energy side (upper panel) are 
bonding and antibonding S $3s$ bands, respectively.
The antibonding S $3p\sigma^*$ band is located 
at about 2.5 eV above $E_{\mathrm F}$. 
The spin split Co $3d$ bands, $t_{2g}$ and $e_g$, are formed between 
S $3p$ and S $3p\sigma^*$ bands (lower panel).
The $e_g$ band near $E_{\mathrm F}$ manifests
nearly half-metallic nature of CoS$_2$.}
\end{figure}
In Fig.~\ref{fig1}, we show the LSDA density of states (DOS) 
obtained at the experimental lattice constant.
Because of strong covalent bonding nature of S-S dimers, each S band 
can be identified with molecular orbitals.
The two peaks in the highest binding energy side 
are the bonding S $3s\sigma$ and 
the antibonding S $3s\sigma^*$ bands, respectively\cite{Bullet}.
The broad band of about 6 eV width between $-4$ and $-10$ eV 
consists mainly of S $3p$, 
which is a mixture of $3p\sigma$, $3p\pi$, and $3p\pi^*$ bands.
The antibonding S $3p\sigma^*$ intra-dimer band is 
located at about $2.5$ eV above the Fermi energy $E_{\mathrm F}$.  
The spin split bands of Co $t_{2g}$ are located between $-1$ and $-3$ eV
below $E_{\mathrm F}$ while Co $e_g$ band are near $E_{\mathrm F}$.   
It is seen that Co $3d$ bands are formed in the energy range
between S $3p$ and S $3p\sigma^*$ bands.
The $t_{2g}$ band is nearly dispersionless with very narrow band width,
whereas the $e_g$ band near $E_{\mathrm F}$ is 
dispersive with relatively large band width of about $2.5$ eV due to 
strong hybridization with S $3p\sigma^*$ band. 
Nearly half-metallic nature and the exchange splitting 
$\Delta_X \sim 1.0$ eV are prominent for $e_g$ band.
  
The present results are slightly different 
from those of existing calculations\cite{Zhao,Yamada}.
In Ref.~\ref{Zhao}, S $3p\sigma^*$ band is at $\sim 1.5$ eV  
above $E_{\mathrm F}$ which is lower than the present result 
by about $1.0$ eV.  Whereas, in Ref.~\ref{Yamada}, 
the states of S $3p\sigma^*$ band is smeared out by
too strong hybridization with Co $e_g$ bands.
The bremsstrahlung isochromat spectrum (BIS) shows 
clear double peak structure at about $1.0$ and $2.5$ eV 
above $E_{\mathrm F}$\cite{Mamiya}, which is 
consistent better with the present result.
That is, the $1.0$ eV peak corresponds to the minority spin $e_g$ 
band and the $2.5$ eV peak to S $3p\sigma^*$ band.   
The occupation of the minority spin $e_g$ band is only $0.04$ electrons/Co,
reflecting that CoS$_2$ is almost half-metallic. 
This feature is in agreement with that obtained by 
Zhao {\it et al.}\cite{Zhao}.
In the analysis of the reflectivity data in terms of
the half-metallic CoS$_2$\cite{Yamamoto},
the minority spin $e_g$ band is assigned at about 1.5 eV 
above $E_{\mathrm F}$. This is close to but a bit higher 
than the present LSDA result.

\begin{figure}[t] 
\epsfig{file=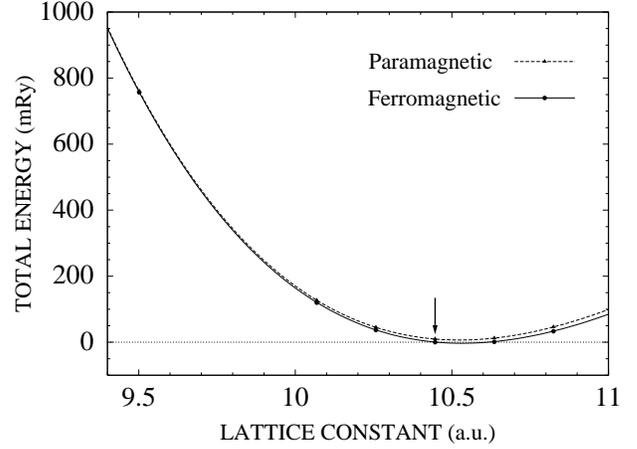,width=8.65cm}
\caption{\label{fig2} The LSDA total energy as a function of lattice constant.
The arrow indicates the experimental lattice constant.
The ferromagnetic ground state is lower in energy than the paramagnetic state 
far below the experimental lattice constant.
}
\end{figure}
Figure~\ref{fig2} is the LSDA total energy as a function of the 
lattice constant.
The ferromagnetic ground state is lower in energy than the paramagnetic state
even far below the experimental lattice constant of $a = 5.528$ \AA~ 
represented by the arrow.
The minimum of the total energy is located near the experimental
lattice constant for both the paramagnetic and the ferromagnetic states,
and thus the LSDA seems to describes well the cohesive bonding properties
of CoS$_2$.
The ferromagnetic ground states is lower in energy by about 10 mRy 
than the paramagnetic state at the experimental lattice constant.
The energy difference between the ferromagnetic and the paramagnetic states 
decreases with decreasing the lattice constant.
This is contrary to the result of Yamada {\it et al.}\cite{Yamada} in which
the paramagnetic is stable near the experimental
lattice constant and the ferromagnetic state becomes stable 
only for $a > 5.57$ \AA.  As they pointed out, it seems to be ascribed to 
small number of k-points they used.
On the basis of the total energy curve shown in Fig.~\ref{fig2}, 
the observed metamagnetic transition behavior under the
high magnetic filed, the high external pressure, and the
chemical pressure by substituting Se in place of S 
can be qualitatively understood \cite{Adachi2,Goto}

\begin{figure}[t]
\epsfig{file=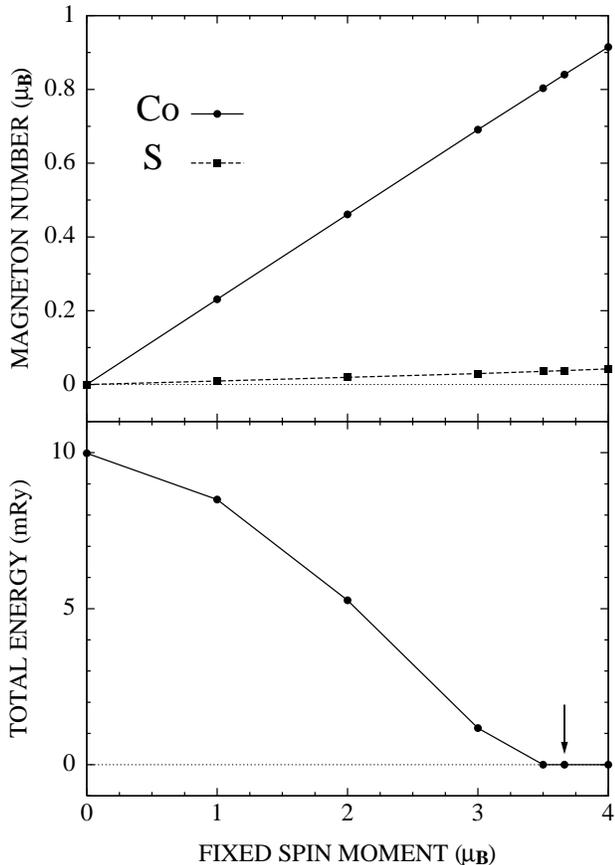,width=8.65cm}
\caption{\label{fig3} The behavior of the spin magnetic moment 
at each site (upper panel) and the total energy (lower panel) 
with varying the total spin magnetic moment $M (\mu_B)$ in the unit cell. 
The arrow denotes the normal-metallic ferromagnetic ground state with
$M = 3.66 \mu_B$.
The half-metallic state with integer $M = 4.00 \mu_B$ is very close in energy 
to the normal-metallic ground state. 
}
\end{figure}
By using the fixed spin moment method within the LSDA,
the total energy is evaluated with varying the magnetic moment in CoS$_2$
(Fig.~\ref{fig3}).
For all the fixed spin moment, $0.00 \mu_B < M \leq 4.00 \mu_B$, 
the ferromagnetic state is stable as compared to the paramagnetic state.
The spin direction of S atoms is always parallel to that of Co.
The ground state is found for $M = 3.66 \mu_B$ (denoted by the arrow) 
which is a normal ferromagnetic metal and lower in energy by about 10 mRy 
than the paramagnetic state. 
It is noticeable that the half-metallic state 
with exact integer magnetic moment of $M = 4.00 \mu_B$ is 
very close in energy to the normal-metallic ground state.
The total energy difference between the normal-metallic ground state 
and the half-metallic state is less than 0.1 mRy
because of quite small electron occupation of the minority spin
$e_g$ band in the ground state.
Thus it is expected that, with only small number of hole doping, 
the half-metallic state can be realized in CoS$_2$\cite{Zhao}.

In general, a polycrystalline half-metallic system is 
expected to show large magnetoresistance (MR) behavior.
It arises from the activated electron tunneling mechanism 
at the grain boundary with the applied external magnetic field
\cite{Hwang}. To our knowledge, there has been no report 
available on the MR measurement for CoS$_2$.
Therefore, the MR measurement on polycrystalline CoS$_2$ 
and Fe$_x$Co$_{1-x}$S$_2$ is desirable to exploit the
correlation between the half-metallic nature and the MR.

\begin{table}[t]
\caption{\label{somm} 
Calculated spin and orbital magnetic moments, $\mu_S$ and $\mu_L$
($\mu_B$) of CoS$_2$ for various Coulomb correlation $U$ (eV).
All the results are from the relativistic calculations.
$M_{fu} (\mu_B)$ is the total magnetic moment per formula unit.
The total magnetic moment per Co atom $\mu_{tot}= 0.86 \mu_B$ 
and the ratio of $\langle L_z \rangle/\langle S_z \rangle = 0.15$
in the LSDA is consistent with experimental data.
}
\begin{tabular}{ccccc}
 $U$    &$M_{fu}$ &      & Co   &  S       \\ \hline 
 0.0    &0.90&$\mu_S$& 0.76 &$~~0.04$  \\ 
        &    &$\mu_L$& 0.06 &$~~0.00$  \\ 
 1.0    &1.06&$\mu_S$& 0.88 &$~~0.04$  \\ 
        &    &$\mu_L$& 0.10 &$~~0.00$  \\ 
 2.0    &1.13&$\mu_S$& 0.93 &$~~0.03$  \\ 
        &    &$\mu_L$& 0.14 &$~~0.00$  \\ 
 3.0    &1.21&$\mu_S$& 1.01 &$~~0.00$  \\ 
        &    &$\mu_L$& 0.20 &$~~0.00$  \\ 
 4.0    &1.25&$\mu_S$& 1.07 &$-0.03$   \\ 
        &    &$\mu_L$& 0.25 &$~~0.00$  \\ \hline
 LSDA   &0.93&$\mu_S$& 0.80 &$~~0.04$  \\  
        &    &$\mu_L$& 0.06 &$~~0.00$  \\
\end{tabular}
\end{table}
We have applied the LSDA and the LSDA+$U$ methods 
to explore the Coulomb correlation effect on the magnetic properties. 
The observed magnetic moments per Co atom
 are in the range of $0.84 \sim 0.91\mu_B$\cite{Adachi1,Yamamoto}.
The orbital magnetic moment in CoS$_2$ is obtained
by including the spin-orbit coupling in the total Hamiltonian. 
In the LSDA + $U$ calculations, we have used fixed $J (=0.89$ eV$)$ 
which is insensitive to local environment. 
In the LSDA, we have obtained the spin and orbital magnetic moments
per Co atom,
$\mu_S = 0.80\mu_B$ and $\mu_L = 0.06\mu_B$, 
respectively (Table~\ref{somm}).
This produces the ratio of 
$\langle L_z \rangle/\langle S_z \rangle = 0.15$, 
which is close to the MCD measurement of 
$\langle L_z \rangle/\langle S_z \rangle = 0.18$\cite{Muro}
and $0.14$\cite{Miyauchi}. 
The total magnetic moment per Co atom $\mu_{tot}= 0.86 \mu_B$ 
in the LSDA is also in agreement with the experiment.

Both the spin and orbital magnetic moments increase with increasing 
the Coulomb correlation parameter $U$. 
For $U \gtrsim 3.0$ eV, the minority spin $e_g$ band moves to higher 
energy states and CoS$_2$ becomes completely half-metallic.
For $U = 4.0$ eV, the minority spin $e_g$ band is formed 
at about 2.0 eV above $E_{\mathrm F}$. 
However, the discrepancy of the total magnetic moment between the
experiment and the theory becomes significant for $U \gtrsim 2.0$ eV.
Moreover, the ratio of $\langle L_z \rangle/\langle S_z \rangle = 0.31$ 
for $U = 2.0$ eV is too large compared 
to the experimental values\cite{Muro,Miyauchi}.   
This finding suggests that the size of the on-site Coulomb correlation in
CoS$_2$ is to be rather small $U \lesssim 1$ eV, and the LSDA gives a 
better description of Co $3d$ states in CoS$_2$.
Note that the LDA results for FeS$_2$ show an excellent agreement 
with PES and BIS data\cite{Folkerts,Imada}. 
In addition, the weak satellite feature in the PES of CoS$_2$\cite{Fujimori}
seems to be consistent with the present results of rather small 
Coulomb correlation $U$.

Finally, let us remark on the electronic structure of CoS$_2$ in 
its paramagnetic phase ($T > T_C$).
As mentioned in the introduction, it is an unresolved issue. 
The question is whether the exchange splitting
in Co $3d$ bands persists above $T_C$ or not.
The present study reveals that the width of Co $e_g$ band is larger than
both the Coulomb correlation interaction $U$ and the exchange splitting
$\Delta_X$.  It thus suggests that the Co $3d$ electrons 
near $E_{\mathrm F}$ will behave as itinerant to manifest Stoner-type 
magnetic phenomena at the finite temperature.  
Indeed, there are such evidences from several experiments.
Resonance PES data for Co $3d$ bands taken at temperatures 
across the magnetic transition
indicate a slight but noticeable spectral change between the paramagnetic
and the ferromagnetic phases, reflecting the long range spin exchange
effect\cite{Muro98}. Optical spectra above $T_C$ are described well
using the itinerant Stoner model\cite{Yamamoto}. 
Also inelastic magnetic neutron scattering data above $T_C$ exhibit the 
Stoner excitations stemming from the itinerant electrons\cite{Hiraka}.
On the other hand, based on the PES studies,
it is claimed that the electronic structure in the paramagnetic phase 
should be described by the local band picture \cite{Fujimori,Mamiya}.
Hence further studies are necessary to ascertain this point.

In conclusion, we have investigated 
the electronic and magnetic properties of the pyrite-type CoS$_2$ 
using the LMTO band method.
We have found the following results:
$(i)$ CoS$_2$ does not belong to the strongly correlated
electron system, because the effect of the Coulomb correlation at Co sites 
is rather small, $U \lesssim 1$ eV,
$(ii)$ CoS$_2$ can be categorized as an itinerant electron ferromagnet,
since the band width of Co $e_g$ state is larger than
$U$ and the exchange splitting $\Delta_X$,
$(iii)$ CoS$_2$ is nearly half-metallic in that the half-metallic
state is very close in energy to the normal-metallic ground state.
A non-negligible orbital magnetic moment of Co $3d$ electrons 
is obtained as $\mu_L = 0.06 \mu_B$ in the LSDA,
which is consistent with experiment.
Accordingly, the electronic and magnetic
properties of CoS$_2$ are described better with the LSDA than
with the LSDA+$U$.

Acknowledgements$-$
The authors would like to thank Jin Ho Park for helpful discussions. 
This work was supported by the KOSEF (1999-2-114-002-5)
and in part by the Korean MOST-FORT fund.

 
\end{document}